\documentclass{article}

\def \g{\gamma}
\def \d{\delta}

\def \et{\eta}

\def \l{\lambda}
\def \m{\mu}
\def \n{\nu}
\def \x{\xi}

\def \r{\rho}

\def \ph{\phi}

\def \La{\Lambda}

\def \Om{\Omega}

\def \la#1{\label{#1}}
\def \ift{\infty}
\def \le{\left}
\def \ri{\right}
\def \da{\dagger}

\def \ti#1{\tilde{#1}}
\def \lb{\lbrack}
\def \rb{\rbrack}

\def \rar{\rightarrow}

\def \ld{\ldots}
\def \cd{\cdots}
\def \nn{\nonumber}
\def \Tr{{\rm Tr} \,}

\newcommand \beq{\begin{eqnarray}}
\newcommand \eeq{\end{eqnarray}}
\newcommand \bea{\begin{eqnarray*}}
\newcommand \eea{\end{eqnarray*}}
\newcommand \ben{\begin{enumerate}}
\newcommand \een{\end{enumerate}}
\newcommand \ba{\begin{array}}
\newcommand \ea{\end{array}}

\newtheorem{theorem}{Theorem}[section]
\newtheorem{lemma}{Lemma}[section]

\begin{document}

\begin{center}
\begin{LARGE}
{\bf On Inhomogeneity of a String Bit Model} \\ 
\vspace{.3cm}
{\bf for Quantum Gravity} 
\end{LARGE}

\vspace{.7cm} 
\begin{large}
{\bf C.-W. H. Lee}\footnote{\large e-mail address: lee@scimail.uwaterloo.ca} and 
{\bf R. B. Mann}\footnote{\large e-mail address: mann@avatar.uwaterloo.ca} 
\end{large}
 
\vspace{.7cm} 
\textit{Department of Physics, Faculty of Science, University of Waterloo, Waterloo, Ontario, Canada, N2L 3G1.} \\
\vspace{.4cm} August 9, 2004 \\
\vspace{.7cm} {\large \textbf{Abstract}}
\end{center}

\noindent 
We study quantum gravitational effect on a two-dimensional open universe with one particle by means of a string bit model.  
We find that matter is necessarily homogeneously distributed if the influence of the particle on the size of the universe is 
optimized.

\vspace{.5cm}

\begin{flushleft}
\end{flushleft}

\pagebreak

\section{Introduction}
\label{s1}

As the quest for a quantum theory of gravity continues, two-dimensional theories continue to play an important role 
\cite{dgz, adj, nakayama, al, dgk, dgk01}. Their simplified setting affords considerably greater mathematical progress, 
offering the promise that the insights gleaned from this approach can be implemented in higher dimensions.

Discrete models of quantum gravity have been of increasing interest in this context in recent years.  In two dimensions such 
models are called Hamiltonian string bit models, in which the spatial metric degree of freedom is discretized by introducing 
a distance cutoff $a_* >0$. The volume of spacetime is then $n a_*$ where $n$ is the (integer) number of links in a slice, 
with equal-time slices taken to be polygonal loops for a closed universe or straight lines with discrete segments for an 
open universe.  Each slice is described by a pure quantum state $\left|n\right\rangle$ with $a_*$ fixed, and the set of such 
states forms an orthogonal basis for the Hilbert space of states. Adjacent links are created/destroyed by 
creation/annihiliation operators, preserving the locality of the model. The continuum limits of two-dimensional Lorentzian 
models of quantum gravity can be obtained from these string bit Hamiltonian models \cite{0108149}.
	
It has recently been shown that string-bit models can be extended from the pure gravity case to include couplings to matter 
\cite{0307231}. This is done by introducing new creation and annihilation operators that act to create and destroy particles 
that reside on the links. In this sense the particle is represented by a `coloured link'.  By homogeneously superposing the 
coloured link throughout the equal-time slice, a variety of interesting results can be obtained. Single-particle closed and 
open universe models were found to respectively behave like empty open and closed universes. A broad class of 
closed-universe models with indistinguishable bosons were shown to be special cases of the fractional sector model that 
plays the role of a metamodel for this class. The solution to the fractional sector model at the continuum limit was shown 
to have a transition amplitude that is exactly the same as the $sl(2)$ gravity model in Ref.~\cite{0108149}. 

In this paper we generalize the homogeneous model to include situations in which the particle is in an arbitrary 
superposition of locations on a collection of string bits.  In other words, the matter distribution may be inhomogeneous on 
a time slice.  We find that if certain parameters in the model that measure the effect of the particle on the size of a 
time-slice take on specific values, then inhomogeneity will not arise; the model is physically equivalent to homogenous 
single-particle models, having the same transition amplitude in the continuum limit.	

The outline of our paper is as follows.  In Section~\ref{s2}, we briefly review the general formalism of the string bit 
model and describe in detail the model we study in subsequent sections.  In Sections~\ref{s3} and \ref{s4}, we work out the 
eigenstates of this model in asymptotic form.  In Section~\ref{s5}, we classify the solutions.  One class of solutions 
describes two empty open universes glued together; another class describes a universe with homogeneous matter distribution;
yet another class offers other possibilities.  In Section~\ref{s7}, we summarize the key results of this article scattered 
throughout Sections~\ref{s2} to \ref{s5}.  

\section{The model}
\label{s2}

Consider an open two-dimensional universe with two boundaries.  Considering it as a succession of time slices, we discretize 
each time slice into a number of links.  The two outermost links represent the two boundaries, some interior links represent 
empty space, and other interior links represent matter; we will call them boundary links, empty links, and matter links, 
respectively.  For the sake of simplicity, in this article we will consider only the case in which there is exactly one 
indestructible and indivisible particle in a time slice.  Hence there are exactly one matter link, two boundary links and an 
arbitrary non-negative number of empty links.  The matter link may drift as time goes by, and the number of empty links may 
change.

Mathematically, a link is represented by a matrix of creation operators.  In particular, the bosonic creation operators 
$a^{\da \m}_{\n}$, where $\m$ and $\n$ are any positive integers between 1 and a positive integer $N$ inclusive, represent 
an empty link.  In addition, the creation operators 
\[ \bar{q}^{\da \m}, \; q^{\da}_{\m}, \; \mbox{and} \; b^{\da \m}_{\n} \]
represent the left-boundary link, the right-boundary link, and the matter link, respectively.  Because we consider a 
one-particle model, it does not matter whether these three kinds of creation operators are bosonic or 
fermionic\footnote{However whether they are bosonic or fermionic leads to different results in more complicated models.}; 
for the sake of definiteness, we will take all these operators to be bosonic.  The corresponding annihilation operators are
\[ a^{\n}_{\m}, \; \bar{q}_{\m}, \; q^{\m}, \; \mbox{and} \; b^{\n}_{\m}, \]
respectively.  They satisfy the standard canonical commutation relations:
\bea
   \le\lb a^{\m_2}_{\m_1}, a^{\da \m_4}_{\m_3} \ri\rb & = & \d^{\m_4}_{\m_1} \d^{\m_2}_{\m_3}, \\
   \le\lb b^{\m_2}_{\m_1}, b^{\da \m_4}_{\m_3} \ri\rb & = & \d^{\m_4}_{\m_1} \d^{\m_2}_{\m_3}, \\
   \le\lb q^{\m_2}, q^{\da}_{\m_1} \ri\rb & = & \d^{\m_2}_{\m_1}, \; \mbox{and} \\
   \le\lb \bar{q}_{\m_1}, \bar{q}^{\da \m_2} \ri\rb & = & \d^{\m_2}_{\m_1}.
\eea
All other commutators among these operators vanish.

The quantum state of a time slice with $m$ empty links to the left and $n$ empty links to the right of the matter link, 
where $m$ and $n$ are non-negative integers, is
\[ |m, n\rangle := \frac{1}{N^{(m+n+2)/2}} \bar{q}^{\da} (a^{\da})^m b^{\da} (a^{\da})^n q^{\da} |\Om \rangle, \]
where $|\Om \rangle$ is the vacuum state and the matrix product $\bar{q}^{\da} (a^{\da})^m b^{\da} (a^{\da})^n q^{\da}$ is 
defined as
\[ \bar{q}^{\da} (a^{\da})^m b^{\da} (a^{\da})^n q^{\da} := \bar{q}^{\da \m_1} a^{\da \m_2}_{\m_1} a^{\da \m_3}_{\m_2} \cd
   a^{\da \m_{m+1}}_{\m_m} b^{\da \n_1}_{\m_{m+1}} a^{\da \n_2}_{\n_1} a^{\da \n_3}_{\n_2} \cd a^{\da \n_{n+1}}_{\n_n}
   q^{\da}_{\n_{n+1}} \]
where we have employed the summation convention.  
In the large-$N$ limit, the norm of this state is 1:
\[ \lim_{N \rar \ift} \langle m, n|m, n\rangle = 1 \]
The Hilbert space ${\cal H}$ of all quantum states of a time slice is spanned by all possible $|m, n\rangle$.  

A suitable Hamiltonian for this one-particle open-universe model\footnote{The differences between this Hamiltonian 
and the Hamiltonian of a similar model in Ref.~\cite{0307231} are:
\ben
\item the coefficients of $\bar{q}^{\da} \bar{q}$ and $(q^{\da})^t q^t$ are 1/4 instead of -1/4, and
\item $\g = 1$ in the Hamiltonian in Ref.~\cite{0307231}.
\een}  is:
\[ H := H_0 + K H_K + \l H_{-1} + \l H_1, \]
where 
\beq
   H_0 & := & \Tr a^{\da} a + V \Tr b^{\da} b + \frac{1}{4} \bar{q}^{\da} \bar{q} + \frac{1}{4} (q^{\da})^t q^t, 
\la{2.2} \\
   H_K & := & - \frac{1}{N} \Tr a^{\da} b^{\da} ab - \frac{1}{N} \Tr b^{\da} a^{\da} ba  
   - \frac{\g}{N} \bar{q}^{\da} b^{\da} b \bar{q} - \frac{\g}{N} (q^{\da})^t (b^{\da})^t b^t q^t, 
\la{2.3} \\
   H_{-1} & := & \frac{1}{\sqrt{N}} \Tr (a^{\da})^2 a + 
   \frac{\x}{\sqrt{N}} \le( \Tr b^{\da} a^{\da} b + \Tr a^{\da} b^{\da} b \ri) \nn \\
   & & + \frac{\et}{N \sqrt{N}} \le( \bar{q}^{\da} b^{\da} a^{\da} b \bar{q} 
   + (q^{\da})^t (b^{\da})^t (a^{\da})^t b^t q^t \ri),
\la{2.4}
\eeq
and
\beq
   \lefteqn{H_1 := \frac{1}{\sqrt{N}} \Tr a^{\da} a^2} \nn \\
   & & + \frac{\x}{\sqrt{N}} \le( \Tr b^{\da} ab + \Tr b^{\da} ba \ri) 
   + \frac{\et}{N \sqrt{N}} \le( \bar{q}^{\da} b^{\da} a b \bar{q} + (q^{\da})^t (b^{\da})^t a^t b^t q^t \ri). 
\la{2.5}
\eeq
The superscripts $t$ in Eqs.(\ref{2.2}), (\ref{2.3}), (\ref{2.4}), and (\ref{2.5}) denote the transposes of the matrices 
concerned.  

Let us describe the action of $H$ on ${\cal H}$ in the large-$N$ limit.  (See ref.~\cite{9906060} for a review of how any
operator that is the trace of a product of matrices of creation or annihilation operators introduced above behaves in the 
large-$N$ limit.)  In $H_0$, the first term counts the number of empty links in a time-slice, thereby measuring the energy 
of empty space, which is proportional to its size.   The second term may be interpreted as a mass term for the matter with 
$V$ being the mass.  The third and fourth terms are the energy terms of the left and right boundaries, respectively.  $H_K$ 
is a variant of the Hubbard model in the context of the string bit model \cite{9806019}.  The action of the first two terms 
of $H_K$ on $|m, n\rangle$ causes the matter link to respectively drift to the right or left.  The third and fourth terms 
may be thought of as open boundary conditions for this Hubbard model.  $\g$ is a constant.  If $\g = 1$, then $H_K$ leaves 
invariant the subspace of ${\cal H}$ consisting of all quantum states of time slices in which matter is homogeneously 
distributed \cite{0307231}.  The coefficient $K$ in $H$ may be interpreted as the Hubbard constant; the larger its value, 
the easier the matter link drifts.

$H_{-1}$ is a collection of terms whose action increases the number of links of a discretised time slice by one.  The first
term causes an empty link to split into two.  The terms proportional to $\x$, a constant, cause the matter link to split 
into itself and an empty link; the new empty link may reside on the left or right of the matter link.  The terms 
proportional to $\et$, another constant, also cause the matter link to split into itself and an empty link if the matter 
link is adjacent to a boundary link; the new empty link must be on the interior side of the matter link.  $H_1$ is the 
Hermitian conjugate of $H_{-1}$.  The action of any term of $H_1$ causes two adjacent links to combine together.  The 
constant coefficient $\l$ measures the ease with which a link splits into two or the ease with which adjacent links combine 
together.

We see from the above description that $H$ is the simplest non-trivial Hamiltonian that allows for a variation of the 
size of the time slice and the matter link to move around as time goes by.  Furthermore, the matter link may interact with
the boundary links in a special manner as displayed by the boundary terms.  

\section{An Asymptotic Expression for Eigenstates}
\la{s3}

We consider now general superpositions of states $|m, n\rangle$, where
\beq
   \ph := \sum_{m, n=0}^{\ift} a_{mn} |m, n\rangle 
\la{3.0a}
\eeq
is an eigenstate of $H$ and $E$ the corresponding eigenenergy.  If any two coefficients of quantum states with the same 
total number of links are equal, i.e. if 
\[ a_{m + k, n - k} = a_{mn} \]
for any non-negative values of $m$ and $n$ and any negative value of $k$ such that $k \leq n$, then matter is homogeneously 
distributed.  The action of the Hamiltonian on $\ph$ yields
\beq
   \lefteqn{\le( m + n + \frac{1}{2} + V \ri) a_{mn}} \nn \\
   & & - K a_{m+1, n-1} - K a_{m-1, n+1} + \l (m - 1 + \xi) a_{m-1, n} + \l (n - 1 + \xi) a_{m, n-1} \nn \\
   & & + \l (m + \xi) a_{m+1, n} + \l (n + \xi) a_{m, n+1} = E a_{mn}
\la{3.1}
\eeq
for any positive integers $m$ and $n$, 
\beq
   \lefteqn{\le( n + \frac{1}{2} + V - \g K \ri) a_{0n} - K a_{1, n-1}} \nn \\
   & &  + \l (n - 1 + \xi + \et) a_{0, n-1} + \xi \l a_{1n} + \l (n + \xi + \et) a_{0, n+1} = E a_{0n}
\la{3.2}
\eeq
for any positive integer $n$,
\beq
   \lefteqn{\le( m + \frac{1}{2} + V - \g K \ri) a_{m0} - K a_{m-1, 1}} \nn \\
   & & + \l (m - 1 + \xi + \et) a_{m-1, 0} + \xi \l a_{m1} + \l (m + \xi + \et) a_{m+1, 0} = E a_{m0}
\la{3.3}
\eeq
for any positive integer $m$, and
\beq
   \le( \frac{1}{2} + V - 2 \g K \ri) a_{00} + \l (\xi + \et) (a_{10} + a_{01}) = E a_{00}. 
\la{3.4}
\eeq  

It follows from Eq.~(\ref{3.1}) that 
\[ m (\l a_{m+1, n} + a_{mn} + \l a_{m-1, n}) + n (\l a_{m, n+1} + a_{mn} + \l a_{m, n-1}) \simeq 0 \]
if both $m$ and $n$ are large.  Hence
\[ a_{mn} \sim p^{m+n}, \]
where $p$ is the solution to the quadratic equation
\beq
   \l p^2 + p + \l = 0. 
\la{3.4a}
\eeq
It is thus natural to write $a_{mn}$ as
\beq
   a_{mn} := b_{mn} p^{m+n} 
\la{3.5}
\eeq
for any non-negative integers $m$ and $n$.  Normalizability implies that $b_{mn}$ increases with $m$ and $n$ polynomially.
Since we are only interested in the continuum limit at which 
\[ \l = - \frac{1}{2} \exp \le( - \frac{\La a_*^2}{2} \ri), \]
where the parameter $a_*$ tends to 0, and the number of string bits in a time slice is inversely proportional to $a_*$, we 
may try the ansatz
\beq
   b_{mn} := \sum_{r=0}^{\ift} \sum_{s=0}^{\ift} c_{rs} a_*^{r+s} m^r n^s + {\cal O}(a_*),
\la{3.6}
\eeq
where both $m$ and $n$ are of order $1/a_*$ so that $a_*^{r+s} m^r n^s$ is of order unity for any values of $r$ and $s$. The 
$c_{rs}$ are constants and only a finite number of them are non-zero.  Higher order terms in $b_{mn}$ may be written as
\[ \sum_{i=1}^{\ift} \sum_{r=0}^{\ift} \sum_{s=0}^{\ift} c_{rs}^{(i)} a_*^{r+s+i} m^r n^s, \]
where $c_{rs}^{(i)}$ are constants.

Upon substitution of Eqs.~(\ref{3.5}) and (\ref{3.6}) into Eq.~(\ref{3.1}) we find
\beq
   \lefteqn{\l (m + \x) p^2 \sum_{r=0}^{\ift} \sum_{s=0}^{\ift} c_{rs} a_*^{r+s} (m+1)^r n^s} \nn \\ 
   & & + \l (n + \x) p^2 \sum_{r=0}^{\ift} \sum_{s=0}^{\ift} c_{rs} a_*^{r+s} m^r (n+1)^s \nn \\
   & & + \le( m + n + \frac{1}{2} + V - E \ri) p \sum_{r=0}^{\ift} \sum_{s=0}^{\ift} c_{rs} a_*^{r+s} m^r n^s \nn \\
   & & - Kp \sum_{r=0}^{\ift} \sum_{s=0}^{\ift} c_{rs} a_*^{r+s} (m+1)^r (n-1)^s \nn \\
   & & - Kp \sum_{r=0}^{\ift} \sum_{s=0}^{\ift} c_{rs} a_*^{r+s} (m-1)^r (n+1)^s \nn \\
   & & + \l (m - 1 + \x) \sum_{r=0}^{\ift} \sum_{s=0}^{\ift} c_{rs} a_*^{r+s} (m-1)^r n^s \nn \\
   & & + \l (n - 1 + \x) \sum_{s=0}^{\ift} \sum_{s=0}^{\ift} c_{rs} a_*^{r+s} m^r (n-1)^s \simeq 0.
\la{3.7}
\eeq
Using Eq.~(\ref{3.4a}) to simplify Eq.~(\ref{3.7}) reveal that there are no terms of order $1/a_*$ on the left hand side 
of Eq.~(\ref{3.7}).  It also follows from Eq.~(\ref{3.4a}) that terms neglected in the ansatz~(\ref{3.6}) contribute to 
terms of order~$a_*$ or higher on the left hand side of Eq.~(\ref{3.7}).

Assume that the continuum limit is well defined.  Then the eigenenergy $E$ is order $a_*$.  Let us introduce the following
Maclaurin series expansions in $a_*$:
\bea
   V & := & V_0 + V_1 \sqrt{\La} a_* + {\cal O} (a_*^2), \\
   K & := & K_0 + K_1 \sqrt{\La} a_* + {\cal O} (a_*^2), \\
   \g & := & \g_0 + \g_1 \sqrt{\La} a_* + {\cal O} (a_*^2), \\
   \x & := & \x_0 + \x_1 \sqrt{\La} a_* + {\cal O} (a_*^2), \; \mbox{and} \\
   \et & := & \et_0 + \et_1 \sqrt{\La} a_* + {\cal O} (a_*^2). 
\eea
After some algebra, we find the sum of the terms of order unity on the left hand side of Eq.~(\ref{3.7}) to be
\[ \sum_{r=0}^R \sum_{s=0}^S c_{rs} a_*^{r+s} m^r n^s \le( \frac{3}{2} + V_0 - 2 K_0 - 2 \x_0 \ri). \]
Hence
\beq
   \frac{3}{2} + V_0 - 2 K_0 - 2 \x_0 = 0
\la{3.8}
\eeq
is a necessary condition for the existence of the continuum limit.  It follows from Eqs.~(\ref{3.8}) and (\ref{3.4a}) that 
terms neglected in the ansatz~(\ref{3.6}) contribute to terms of order~$a_*^2$ or higher on the left hand side of 
Eq.~(\ref{3.7}).

The sum of the terms of order~$a_*$ on the left hand side of Eq.~(\ref{3.7}) is
\bea
   & - & \frac{1}{2} \sum_{r=0}^{\ift} \sum_{s=0}^{\ift} c_{r+1, s} a_*^{r+s+1} m^r n^s (r + 1)^2 \\
   & - & \frac{1}{2} \sum_{r=0}^{\ift} \sum_{s=0}^{\ift} c_{r, s+1} a_*^{r+s+1} m^r n^s (s + 1)^2 \\
   & + & \sum_{r=0}^{\ift} \sum_{s=0}^{\ift} c_{rs} a_*^{r+s} m^r n^s 
   \le(r + s + V_1 - 2 K_1 - 2 \x_1 + 1 - \frac{E}{\sqrt{\La} a_*} \ri) \sqrt{\La} a_*.
\eea
Hence
\beq
   \lefteqn{c_{rs} \sqrt{\La} \le( r + s + V_1 - 2 K_1 - 2 \x_1 + 1 - \frac{E}{\sqrt{\La} a_*} \ri)} \nn \\
   & & - \frac{1}{2} c_{r+1, s} (r+1)^2 - \frac{1}{2} c_{r, s+1} (s+1)^2 = 0 
\la{3.9}
\eeq
for all non-negative integer values of $r$ and $s$.  

We solve this relation by assuming that the series in $c_{rs}$ truncates. Let $R_0$ and $S_0$ be non-negative integers such 
that $c_{R_0 S_0} \neq 0$ but $c_{R_0 + r', S_0 + s'} = 0$ for any non-negative integers $r'$ and $s'$ such that $r' + s' > 
0$.  Then Eq.~(\ref{3.9}) implies 
that
\beq
   E = (1 + Q + \ti{h}) \sqrt{\La} a_*, 
\la{3.10}
\eeq
where $Q = R_0 + S_0$ and $\ti{h} = V_1 - 2 K_1 - 2 \x_1$.  Thus if $R'$ and $S'$ are non-negative integers such that 
$c_{R'S'} \neq 0$ but $c_{R' + r', S' + s'} = 0$ for any non-negative integers $r'$ and $s'$ such that $r' + s' > 0$, then 
$R' + S' = R_0 + S_0$.  One may then readily verify that 
\beq
   c_{rs} = \sum_{R=0}^Q (-2 \sqrt{\La})^{r + s - Q} \frac{R! (Q - R)!}{r! s!} 
   \le( \ba{c} R \\ r \ea \ri) \le( \ba{c} Q - R \\ s \ea \ri) c_{R, Q-R},
\la{3.11}
\eeq
is the unique solution to Eq.~(\ref{3.9}).  Combining Eqs.~(\ref{3.0a}), (\ref{3.5}), (\ref{3.6}) and (\ref{3.11}) yields an 
asymptotic expression for any eigenstate.  We will determine the constraints on the values of $Q$ and $c_{R, Q-R}$ in the 
next section.

\section{Boundary Conditions}
\la{s4}

Let us analyze Eqs.~(\ref{3.2}), (\ref{3.3}), and (\ref{3.4}) in this section.  Let
\[ f_{Q,R} := \frac{R! (Q - R)!}{(-2 \sqrt{\La})^Q} c_{Q-R, R}. \]
It then follows from Eqs.~(\ref{3.6}) and (\ref{3.11}) that
\beq
   b_{mn} \simeq \sum_{S = 0}^Q f_{QS} \sum_{r=0}^{Q-S} \sum_{s=0}^S \le( \ba{c} Q - S \\ r \ea \ri) 
   \le( \ba{c} S \\ s \ea \ri) \frac{(-2 \sqrt{\La} a_* m)^r (-2 \sqrt{\La} a_* n)^s}{r! s!}.
\la{4.1}
\eeq
Substituting Eqs.~(\ref{3.5}) and (\ref{4.1}) into Eq.~(\ref{3.2}) yields
\beq
   \lefteqn{\le( n + \frac{1}{2} + V - \g K - E \ri) p \sum_{S=0}^Q f_{QS} \sum_{s=0}^S \frac{(-2 \sqrt{\La} a_*)^s}{s!}
   \le( \ba{c} S \\ s \ea \ri) n^s} \nn \\
   & & - Kp \sum_{S=0}^Q f_{QS} \sum_{r=0}^{Q-S} \sum_{s=0}^S \frac{(-2 \sqrt{\La} a_*)^{r+s}}{r! s!}
   \le( \ba{c} Q - S \\ r \ea \ri) \le( \ba{c} S \\ s \ea \ri) (n-1)^s \nn \\
   & & + \l (n - 1 + \x + \et) \sum_{S=0}^Q f_{QS} \sum_{s=0}^S \frac{(-2 \sqrt{\La} a_*)^s}{s!}
   \le( \ba{c} S \\ s \ea \ri) (n-1)^s \nn \\
   & & + \l \xi p^2 \sum_{S=0}^Q f_{QS} \sum_{r=0}^{Q-S} \sum_{s=0}^S \frac{(-2 \sqrt{\La} a_*)^{r+s}}{r! s!}
   \le( \ba{c} Q - S \\ r \ea \ri) \le( \ba{c} S \\ s \ea \ri) n^s \nn \\
   & & + \l (n + \x + \et) p^2 \sum_{S=0}^Q f_{QS} \sum_{s=0}^S \frac{(-2 \sqrt{\La} a_*)^s}{s!}
   \le( \ba{c} S \\ s \ea \ri) (n+1)^s \simeq 0
\la{4.2}
\eeq

Once again, we can use Eq.~(\ref{3.4a}) to show that the sum of the terms of order $1/a_*$ on the left hand side of 
Eq.~(\ref{4.2}) vanishes and that terms neglected in the ansatz~(\ref{3.6}) contribute to terms of order~$a_*$ or higher.  
The sum of the terms of order unity is
\[ \sum_{S=0}^Q f_{QS} \sum_{s=0}^S \le( \ba{c} S \\ s \ea \ri) \frac{(-2 \sqrt{\La} a_* n)^s}{s!}
   \le\lb (1 - \g_0) K_0 + \frac{1}{2} \x_0 - \et_0 - \frac{1}{2} \ri\rb. \]
where we used Eq.~(\ref{3.8}) and the various Maclaurin series.  Hence
\beq
   (1 - \g_0) K_0 + \frac{1}{2} \x_0  - \et_0 - \frac{1}{2} = 0
\la{4.3}
\eeq
is another necessary condition for the continuum limit to be well defined.  It follows from Eqs.~(\ref{4.3}) and 
(\ref{3.4a}) that terms neglected in the ansatz~(\ref{3.6}) contribute to terms of order~$a_*^2$ or higher on the left hand 
side of Eq.~(\ref{4.2}).

The sum of the terms of order $a_*$ on the left hand side of Eq.~(\ref{4.2}) is
\bea
   \lefteqn{\sum_{S=0}^Q f_{QS} \sum_{s=0}^S \le( \ba{c} S \\ s \ea \ri) \frac{(-2 \sqrt{\La} a_* n)^s}{s!} \sqrt{\La} a_* 
   \le\lb (\x_0 + 2 K_0 - 1) (Q - S) \ri. } \\
   && \le. - 2K_0 \frac{S - s}{s + 1} + (-1 + \g_0 + \g_1) K_0 + (1 - \g_0) K_1 - \frac{7}{2} \x_1 + \et_0 - \et_1 \ri\rb.
\eea
(We have made use of Eq.~(\ref{3.8}) to obtain the above formula.)  Since $n$ is arbitrary, we obtain the constraint
\beq 
   \sum_{S = s}^Q f_{QS} \le( \ba{c} S \\ s \ea \ri) \le\lb (\x_0 - 1) (Q - S) 
   + 2K_0 \le( Q - S \frac{s+2}{s+1} + \frac{s}{s+1} \ri) + C \ri\rb = 0,
\la{4.5a}
\eeq
where
\[ C := (-1 + \g_0 + \g_1) K_0 + (1 - \g_0) K_1 - \frac{7}{2} \x_1 + \et_0 - \et_1 \]
is a constant and $s$ is any non-negative integer.

Repeating the above argument for Eq.~(\ref{3.3}) leads to the same necessary Constraint~(\ref{4.3}) and the constraint
\beq
   \lefteqn{\sum_{R = r}^Q f_{Q, Q-R} \le( \ba{c} R \\ r \ea \ri) \le\lb (\x_0 - 1) (Q - R) \ri. } \nn \\
   && \le. + 2K_0 \le( Q - R \frac{r+2}{r+1} + \frac{r}{r+1} \ri) + C \ri\rb = 0
\la{4.6}
\eeq
for any non-negative value of $r$.

Let us turn our attention to Eq.~(\ref{3.4}). Substituting Eqs.~(\ref{3.5}) and (\ref{4.1}) into Eq.~(\ref{3.4}) and using 
(\ref{3.8}) and (\ref{4.3}), we get
\bea
   \lefteqn{\et_0 \sum_{S = 0}^Q f_{QS} + \sqrt{\La} a_* \lb (\x_0 + \et_0 - 1) (Q + 1)} \\
   && + 2(1 - \g_0) K_1 - 2 \g_1 K_0 + \x_1 - \et_1 \rb \sum_{S = 0}^Q f_{QS} + \r(a_*) = 0, 
\eea
where $\r(a_*)$ comes from terms neglected in the ansatz~(\ref{3.6}).  It is of order~$a_*$ or higher and is directly 
proportional to $\et_0$.  As a result, 
\beq
   \et_0 \sum_{S = 0}^Q f_{QS} & = & 0, 
\la{4.7} 
\eeq
and
\beq
   \lefteqn{\lb (\x_0 + \et_0 - 1) (Q + 1)} \nn \\
   && + 2 (1 - \g_0) K_1 - 2 \g_1 K_0 + \x_1 - \et_1 \rb \sum_{S = 0}^Q f_{QS} + \frac{\r (a_*)}{\sqrt{\La} a_*} = 0
\la{4.8}
\eeq
are two additional necessary conditions for the continuum limit to be well defined.

\section{Different Scenarios}
\la{s5}

Let us solve the constraints and classify the solutions.

\subsection{Direct Product of Empty Open Universes}

One solution to Constraints~(\ref{4.7}) and (\ref{4.8}) is $\et_0 = 0$ (which in turn implies that $\r(a_*)=0$), $\x_0 = 1$,
and
\beq 
   2 (1 - \g_0) K_1 - 2 \g_1 K_0 + \x_1 - \et_1 = 0.
\la{5.0}
\eeq
Suppose that the matter link cannot drift at the continuum limit, i.e.
\[ K_0 = 0. \]
Then Constraints~(\ref{4.5a}) and (\ref{4.6}) lead to
\[ C = 0, \]
and the ratios $f_{Q0}: f_{Q1}: \cd :f_{QQ}$ is arbitrary.  Furthermore, Eq.~(\ref{3.8}) implies that $V_0 = 1/2$.  
Comparison of this value of $V_0$ with the coefficients of the boundary terms in Eq.~(\ref{2.2}) reveals that the matter 
link is nothing but the connection of a right and a left boundary link.  Moreover, comparison of the energy spectra of this 
one-particle model (\ref{3.10}) and the $sl(2)$ gravity model with $h = 1/2$ (Eq.~(30) of 
Ref.~\cite{0108149})\footnote{Note that $\ti{h} = 0$ in Ref.~\cite{0108149}.}, and comparison of the asymptotic expressions 
of the eigenstates of this model (Eqs.~(\ref{3.0a}), (\ref{3.5}), and (\ref{4.1})) and of the $sl(2)$ gravity model as 
partly displayed in Eq.~(31) of Ref.~\cite{0108149} confirm that this model is equivalent to two empty open universes joined 
together at one of their ends.

The transition amplitude at the continuum limit is defined as
\beq
   \ti{G} (L_1, L_2; L'_1, L'_2; T) := \lim_{a_* \rar 0} \frac{1}{a_*^2} \langle \frac{L'_1}{a_*}, \frac{L'_2}{a_*} 
   |e^{-tH}| \frac{L_1}{a_*}, \frac{L_2}{a_*} \rangle.
\la{5.7}
\eeq
It follows immediately from the argument in the preceding paragraph and Eq.~(38) of Ref.~\cite{0108149} that
\beq
   \lefteqn{\ti{G} (L_1, L_2; L'_1, L'_2; T) = \frac{\La e^{-2 \ti{h} \sqrt{\La} T}}{\sinh^2 (\sqrt{\La} T)}} \nn \\ 
   && \cdot e^{- \sqrt{\La} (L_1 + L_2 + L'_1 + L'_2) \coth (\sqrt{\La} T)} 
   I_0 \le( \frac{2 \sqrt{\La L_1 L'_1}}{\sinh (\sqrt{\La} T)} \ri)
   I_0 \le( \frac{2 \sqrt{\La L_2 L'_2}}{\sinh (\sqrt{\La} T)} \ri).
\la{5.8} 
\eeq  

\subsection{Homogeneous Matter Distribution}

Consider again $\et_0 = 0$, $\x_0 = 1$, and $2 (1 - \g_0) K_1 - 2 \g_1 K_0 + \x_1 - \et_1 = 0$ (Eq.~(\ref{5.0})).  Assume, 
however, that $K_0 \neq 0$ so that the matter link may drift.  We will show by contradiction that 

\begin{theorem}
\[ C = 0. \]  
\end{theorem}
{\bf Proof}.  Suppose, on the contrary, that $C \neq 0$.  Then Eq.~(\ref{4.5a}) with $s = Q$ implies that
\[ f_{QQ} = 0. \]
Similarly, Eq.~(\ref{4.6}) with $r = Q$ implies that
\beq
   f_{Q0} = 0.
\la{5.11}
\eeq
Treat Eq.~(\ref{4.5a}) with $s$ = 1, 2, \ld, $Q - 1$ as a set of simultaneous linear equations in $f_{Q1}$, $f_{Q2}$, \ld, 
and $f_{Q, Q-1}$.  A non-trivial solution exists only if the determinant of a matrix of coefficients vanishes, i.e.
\[ \prod_{s = 1}^{Q-1} \le( \ba{c} s \\ s \ea \ri) 
   \le\lb 2 K_0 \le( Q - s \frac{s+2}{s+1} + \frac{s}{s+1} \ri) + C \ri\rb = 0. \]
Hence there exists a positive integer $s_0 < Q$ such that 
\[ 2 K_0 \le( Q - s_0 \frac{s_0 + 2}{s_0 + 1} + \frac{s_0}{s_0 + 1} \ri) + C = 0. \]
Thus
\[ C = - 2 K_0 (Q - s_0). \]

It then follows from Eq.~(\ref{4.5a}) with $s$ = $Q-1$, $Q-2$, \ld, and $s_0 + 1$ that
\[ f_{Q, Q-1} = f_{Q, Q-2} = \cd = f_{Q, s_0 + 1} = 0 \]
and from the same equation that
\bea
   \sum_{S = s}^{s_0} f_{Qs} \le( \ba{c} S \\ s \ea \ri) \le( s_0 - S \frac{s+2}{s+1} + \frac{s}{s+1} \ri) = 0
\eea
for $s$ = 0, 1, \ld, and $s_0 - 1$.  Theorem~\ref{t2} in Appendix~\ref{sa1} with $Q' = s_0$ and Eq.~(\ref{5.11}) then imply 
that
\[ f_{Q, s_0} = f_{Q, s_0 - 1} = \cd = f_{Q0} = 0. \]
However, this contradicts the assumption that there is a non-trivial solution among $f_{Q1}$, $f_{Q2}$, \ld, and 
$f_{Q, Q-1}$.  Q.E.D.

\vspace{1em}

It follows from the above theorem and Eq.~(\ref{4.5a}) that $f_{Q0}$, $f_{Q1}$, \ld, and $f_{QQ}$ must satisfy
\beq
   \sum_{S = s}^Q f_{QS} \le( \ba{c} S \\ s \ea \ri) \le( Q - S \frac{s+2}{s+1} + \frac{s}{s+1} \ri) = 0 
\la{4.5}
\eeq
for any non-negative value of $s$.  Consequently, Theorem~\ref{t2} in Appendix~\ref{sa1} with $Q' = Q$ implies that 
\[ f_{Q0} = f_{Q1} = \cd = f_{QQ} \]
is the only solution to Eqs.~(\ref{4.5a}) and (\ref{4.6}) if $\x_0 = 1$.  Then Eq.~(\ref{4.1}) implies that
\beq
   b_{mn} & = & f_{Q0} \sum_{r=0}^Q \sum_{s=0}^{Q-r} \frac{(-2 \sqrt{\La} a_* m)^r (-2 \sqrt{\La} a_* n)^s}{r! s!}
   \sum_{R=0}^Q \le( \ba{c} R \\ r \ea \ri) \le( \ba{c} Q - R \\ s \ea \ri) \nn \\
   & = & f_{Q0} \sum_{r=0}^Q \sum_{s=0}^{Q-r} \frac{(-2 \sqrt{\La} a_* m)^r (-2 \sqrt{\La} a_* n)^s}{r! s!}
   \le( \ba{c} Q + 1 \\ r + s + 1 \ea \ri) \nn \\
   & = & f_{Q0} \sum_{r=0}^Q \sum_{s=r}^Q \frac{(-2 \sqrt{\La} a_* m)^r (-2 \sqrt{\La} a_* n)^{s-r} (Q+1)!}
   {r! (s-r)! (s+1)! (Q-s)!} \nn \\
   & = & f_{Q0} (Q + 1) \sum_{s=0}^Q \sum_{r=0}^s \frac{(-2 \sqrt{\La} a_* m)^r (-2 \sqrt{\La} a_* n)^{s-r}}{(s+1)!}
   \le( \ba{c} Q \\ s \ea \ri) \le( \ba{c} s \\ r \ea \ri) \nn \\
   & = & f_{Q0} (Q + 1) \sum_{s=0}^Q \le( \ba{c} Q \\ s \ea \ri) \frac{(-2 \sqrt{\La} a_*)^s (m + n)^s}{(s+1)!},
\la{5.4a}
\eeq
where the second line follows from a standard combinatoric formula, which may be easily proved by induction on $Q$.  Thus 
$b_{mn}$ and hence $a_{mn}$ depends on $m$ and $n$ via $m+n$.  Physically, this means that on a time slice in any 
eigenstate, the matter link may be found anywhere with the same probability; matter is necessarily homogeneously 
distributed.  It follows from Subsection~5.2 of Ref.~\cite{0108149}, in particular Eq.~(33) of the reference, that 
\beq
   f_{Q0} (Q+1) = 2 \sqrt{(Q+1) \La} a_*. 
\la{5.4b}
\eeq
Substitution of Eqs.~(\ref{3.0a}), (\ref{3.5}), (\ref{5.4a}), (\ref{5.4b}), and (\ref{3.10}) into Definition~(\ref{5.7}) for 
the transition amplitude yields
\bea
   \lefteqn{\ti{G} (L_1, L_2; L'_1, L'_2; T) = 4 \La \sum_{R=1}^{\ift} (R + 1)} \\
   && \cdot \sum_{r=0}^R \sum_{s=0}^R \frac{(-2 \sqrt{\La})^{r+s} (L_1 + L_2)^r (L'_1 + L'_2)^s}{(r+1)! (s+1)!} \\
   && \cdot \le( \ba{c} R \\ r \ea \ri) \le( \ba{c} R \\ s \ea \ri) e^{-\sqrt{\La} (L_1 + L_2 + L'_1 + L'_2)}
   e^{-2 (1 + \ti{h} + R) \sqrt{\La} T}.
\eea
This formula differs from Eq.~(36) of Ref.~\cite{0108149} in which $L = L_1 + L_2$, $L' = L'_1 + L'_2$, and $h = 1$ by a 
factor of
\[ \frac{e^{-2 \ti{h} \sqrt{\La} T}}{L L'} \]
only.  Hence the subsequent calculation is parallel to the one in Ref.~\cite{0108149}, and we may write down the final
expression immediately:
\beq
   \lefteqn{\ti{G} (L_1, L_2; L'_1, L'_2; T) = 
   \sqrt{\frac{\La}{(L_1 + L_2)(L'_1 + L'_2)}} \frac{e^{-2 \ti{h} \sqrt{\La} T}}{\sinh (\sqrt{\La} T)}} \nn \\
   && \cdot e^{- \sqrt{\La} (L_1 + L_2 + L'_1 + L'_2) \coth (\sqrt{\La} T)} 
   I_1 \le( \frac{2 \sqrt{\La (L_1 + L_2) (L'_1 + L'_2)}}{\sinh (\sqrt{\La} T)} \ri).
\la{5.6} 
\eeq

\subsection{Other Possilibities}

If $\et_0 = 0$ but $\x_0 \neq 1$ or Constraint~(\ref{5.0}) does not hold, then Constraint~(\ref{4.8}) implies that
\[ \sum_{S = 0}^Q f_{QS} = 0 \]
for all but perhaps one value of $Q$. If $\et_0 \neq 0$, then Constraint~(\ref{4.7}) implies that 
\[ \sum_{S = 0}^Q f_{QS} = 0 \]
for all values of $Q$.  Furthermore, any solution, if it exists, has to satisfy Constraints~(\ref{4.5a}) and (\ref{4.6}).  
Working out solutions that satisfy all of these constraints remains an interesting research problem.

\section{Conclusion}
\la{s7}

In the context of string bit models, quantum gravitational effects furnish considerable constraints on matter distributions.  
Indeed, the continuum limit for a string bit model with a single particle of matter is well defined only if the constraints
\[ \frac{3}{2} + V_0 - 2 K_0 - 2 \x_0 = 0 \]
(Eq.~(\ref{3.8})), 
\[ (1 - \g_0) K_0 + \frac{1}{2} \x_0  - \et_0 - \frac{1}{2} = 0, \]
(Eq.~(\ref{4.3})), Eq.~(\ref{4.5a}), Eq.~(\ref{4.6}) and Eq.~(\ref{4.8}) are satisfied.  

If $\et = {\cal O} (a_*)$, $\x = 1 + {\cal O} (a_*)$, and Constraint~(\ref{5.0}) holds, we may rewrite Eqs.~(\ref{3.8}) and 
(\ref{4.3}) as $V - 2K - 1/2 = {\cal O} (a_*)$ and $(1 - \g) K = {\cal O} (a_*)$, respectively.  In addition, if $K = 
{\cal O} (a_*)$, then the model describes two empty open universes glued together at one of their ends, and the transition 
amplitude is given by Eq.~(\ref{5.8}); otherwise, matter is necessarily homogeneously distributed, and the transition 
amplitude is given by Eq.~(\ref{5.6}). 

According to Ref.~\cite{0307231}, if we restrict ourselves to the case in which the actions of $H_0 + K H_K$, $H_{-1}$, and
$H_1$ are invariant on the Hilbert space individually, then $\g = 1 + {\cal O} (a_*)$, $\x = 1 + {\cal O} (a_*)$, and 
$\et = {\cal O} (a_*)$.  These are consistent with the findings of the previous paragraph.  We also found that $V - 2K = 3/2 
+ {\cal O} (a_*)$.  At first glance, this violates Constraint~(\ref{3.8}); however, this apparent inconsistency would 
disappear if the coefficients of $\bar{q}^{\da} \bar{q}$ and $(q^{\da})^t q^t$ in $H$ were chosen to be 1/4 instead of -1/4 
as in Ref.~\cite{0307231} (c.f. Eq.~(\ref{2.2})).  This confirms that the model studied in Section~3 of Ref.~\cite{0307231} 
is indeed just a special case of the model we are studying in this article.

The case $\et \neq {\cal O} (a_*)$, $\x \neq 1 + {\cal O} (a_*)$, or Constraint~(\ref{5.0}) does not hold offers other 
possibilities.  Exploration of these possibilities remains an interesting project for the future.

\vskip 1pc \noindent {\Large \textbf{Acknowledgment} \vskip 1pc}

\noindent 
We thank J. Ambj{\o}rn for discussion.  This work was supported in part by the Natural Sciences and Engineering Research
Council of Canada.

\vskip 1pc \noindent {\Large \textbf{Appendix} \vskip 1pc}

\appendix

\section{A Combinatoric Formula}
\la{sa1}

\begin{lemma}
\beq
   \sum_{S = s}^{Q'} \le( \ba{c} S \\ s \ea \ri) \le( Q' - S \frac{s + 2}{s + 1} + \frac{s}{s + 1} \ri) = 0
\la{5.2}
\eeq
for any non-negative value of $s$ and any integer value of $Q'$ such that $Q' \geq s$.
\la{l1}
\end{lemma}
{\bf Proof}.  For a fixed value of $s$, by an inductive argument on $Q'$, one may readily show that 
\beq
   \sum_{n=0}^{N - s} \le( \ba{c} n + s \\ n \ea \ri) = \le( \ba{c} N + 1 \\ s + 1 \ea \ri). 
\la{5.3}
\eeq
It is straightforward to show that Eq.~(\ref{5.2}) holds true for $Q' = s$.  Assume that Eq.~(\ref{5.2}) holds true for
$Q' = Q_0$ for some integer $Q_0$ not less than $s$.  Then
\bea
   & & \sum_{S = s}^{Q_0 + 1} \le( \ba{c} S \\ s \ea \ri) \le( Q_0 + 1 - S \frac{s + 2}{s + 1} + \frac{s}{s + 1} \ri) \\
   & = & \sum_{S = s}^{Q_0} \le( \ba{c} S \\ s \ea \ri) + \le( \ba{c} Q_0 + 1 \\ s \ea \ri) \le\lb Q_0 + 1 
   - (Q_0 + 1) \frac{s + 2}{s + 1} + \frac{s}{s + 1} \ri\rb \\
   & = & \sum_{n=0}^{Q_0 - s} \le( \ba{c} n + s \\ n \ea \ri) - \le( \ba{c} Q_0 + 1 \\ s + 1 \ea \ri) \\
   & = & 0, 
\eea
where the second line follows from the inductive hypothesis and the last line follows from Eq.~(\ref{5.3}).  Hence 
Eq.(\ref{5.2}) holds true for $Q' = Q_0 + 1$ as well.  By mathematical induction,  Eq.~(\ref{5.2}) holds true for all integer 
values of $Q'$ such that $Q' \geq s$.  Q.E.D.

\begin{theorem}
If
\beq
   \sum_{S = s}^{Q'} f_{Q' S} \le( \ba{c} S \\ s \ea \ri) \le( Q' - S \frac{s + 2}{s + 1} + \frac{s}{s + 1} \ri) = 0,
\la{a.1}
\eeq
for $s$ = 0, 1, \ld, and $Q'$.  Then
\[ f_{Q' 0} = f_{Q' 1} = \cd = f_{Q' Q'}. \]
\la{t2}
\end{theorem}
{\bf Proof}.  We may use Eq.~(\ref{a.1}) with $s = Q' - 1$ to determine $f_{Q', Q'-1}$ uniquely in terms of $f_{Q', Q'}$; 
the same equation with $s = Q' - 2$ to determine $f_{Q', Q'-2}$ uniquely in terms of $f_{Q', Q'}$ and $f_{Q', Q'-1}$; \ld; 
and the same equation with $s = 0$ to determine $f_{Q' 0}$ uniquely in terms of $f_{Q'Q'}$, $f_{Q', Q' - 1}$, \ld, and 
$f_{Q' 1}$.  In other words, $f_{Q'0}$, $f_{Q'1}$, \ld, and $f_{Q', Q' - 1}$ depend on $f_{Q'Q'}$ uniquely.  By 
Lemma~\ref{l1},
\[ f_{Q'0} = f_{Q'1} = \cd = f_{Q'Q'} \]
Q.E.D.

\end{document}